\documentclass[preprint,12pt]{elsarticle}

\usepackage{amsmath,amssymb}
\usepackage{graphicx}
\usepackage{booktabs}
\usepackage{tabularx}
\usepackage{makecell}
\usepackage{subcaption}
\usepackage{xspace}
\usepackage{url}
\usepackage{tikz}
\usetikzlibrary{positioning, shapes.geometric, arrows.meta, fit, backgrounds, calc}

\usepackage{microtype}
\usepackage[colorlinks=true, linkcolor=blue, citecolor=blue, urlcolor=blue]{hyperref}

\journal{[Target Journal]}

\begin{document}
	
	\begin{frontmatter}
		
		\title{PTMTD: A Topology-Aware Temporal-Graph Attention Network for Joint Cyber-Attack Detection and Relay-Level Localization in Power System Protection Networks}
		
		\author[1]{Shahin Razani\corref{cor1}\fnref{fn1}}
		\ead{s.razani@email.kntu.ac.ir}
		
		\author[2]{Mohammad Ali Alirezapouri\fnref{fn2}}
		\ead{maa_pouri@mut.ac.ir}
		
		\cortext[cor1]{Corresponding author}
		\fntext[fn1]{Faculty of Electrical Engineering, K.N. Toosi University of Technology, Tehran, Tehran, Iran}
		\fntext[fn2]{Malek Ashtar University of Technology, Department of Electrical Engineering, Tehran, Iran.}
		
		\begin{abstract}
			The digitization of protection relays exposes power grids to sophisticated cyberattacks, yet existing GNN detectors omit temporal encoding, rely on static topologies, and lack relay localization. We propose PTMTD---a Pearson-driven Temporal Multimodal Topology-aware Detection Network---coupling a dilated TCN with multi-head GATv2 in a dual-head architecture for simultaneous 37-class detection and four-relay localization. PTMTD constructs its graph from measurement correlations (no impedance data) and is trained with a composite loss jointly optimizing both tasks. On the MSU--ORNL benchmark (78,377 samples, 37 classes), it achieves 98.70\% accuracy, 98.63\% macro F1, and 95.93\% average relay-wise localization accuracy (measured against Z-score-derived labels), outperforming GraphKAN, LSTM, GCN, and Random Forest by 5.07--10.53 percentage points ($p<0.05$). Ablations show that removing temporal encoding, graph attention, residual connections, or Z-score-derived localization supervision lowers accuracy by 6.56, 6.87, 6.25, and 5.58 percentage points, respectively. With 639k parameters and 2.5~ms CPU inference, PTMTD is lightweight and deployment-ready. The architecture operates only on relay measurements and may therefore transfer to microgrids with similar relay structures, but it has been validated here solely on a transmission-level testbed.
		\end{abstract}
		
		
	\end{frontmatter}
	
	\section{Introduction}
	\label{sec:introduction}
	
	Modern power system protection infrastructure -- spanning transmission-level relay networks, substation automation, and, increasingly, distribution-level microgrids with distributed energy resources (DER) -- has become a cornerstone of grid reliability precisely because it is now pervasively digitized~\cite{tan2025developments}. Central controllers, protective relays, and, in DER-rich settings, inverter-based resources are coordinated through networked sensing, communication, and control layers rather than through isolated, hard-wired logic. This cyber-physical coupling exposes protection systems to an expanding spectrum of adversarial threats, ranging from false data injection (FDI) and denial-of-service (DoS) attacks to stealthy, coordinated manipulations deliberately engineered to evade conventional protection logic~\cite{ali2025cyber}. The 2015 attack on the Ukrainian power grid, which left 225{,}000 customers without electricity, remains a stark demonstration of the physical consequences that can follow from a cyber intrusion, and the underlying risk has only intensified in the years since: cybercrime activity directed at industrial control infrastructure has grown sharply, and cyber threats are now consistently ranked among the most significant operational risks facing critical-infrastructure operators worldwide~\cite{BARBHAYA2025109278}. Protective relay networks -- whether embedded in a transmission-level substation or a distribution-level microgrid -- are especially exposed within this broader threat landscape, owing to insecure legacy communication protocols, remote and lightly monitored deployment sites, and a tight coupling between control loops and physical dynamics that allows a single compromised measurement or command to propagate rapidly into cascading equipment faults, voltage instability, or full-scale outages~\cite{tan2025developments,saqib}. Against this backdrop, the development of intrusion detection systems (IDS) that are simultaneously accurate, interpretable, domain-knowledge grounded, and computationally lightweight enough for real-time deployment on resource-constrained relay and controller hardware has become a pressing research priority.
	
	The literature on cyber-attack detection spans a wide methodological spectrum, from classical signal-processing and observer-based schemes -- often developed for microgrid settings specifically -- to modern deep learning and, most recently, graph-structured representations of the cyber-physical system. We survey this landscape in detail in Section~\ref{sec:related_work}; here we summarize the central gap that motivates the present work. Observer- and control-theoretic approaches~\cite{razani2025indirect,razani2026robust,hosseini2026cyber,11168950} offer strong physical interpretability and, in many cases, provable stability guarantees, but they are typically formulated around low-dimensional, single-plant dynamics and do not scale naturally to the fine-grained, many-class attack taxonomies encountered in realistic substation and relay networks. Statistical and signal-transform methods, often paired with blockchain or watermarking mechanisms~\cite{SUBRAMANIAM2025103400,ghiasi2021cyber,BALAHEWA2026113148}, provide lightweight, physically motivated detection but remain largely confined to binary or coarse-grained attack identification. Deep learning frameworks built on LSTM, autoencoder, convolutional, and hybrid architectures~\cite{ali2025cyber,saqib,fakhrooeian2025artificial,11098679,eswaran2025cyberattack,orman2025cyberattack,BARBHAYA2025109278,ogiesobaeguakun2026cyberattackdetectionvirtualizedmicrogrids,ZHENG2026112529,RAGAB2025438} achieve high raw detection accuracy but generally treat the system as a flat feature vector or a collection of independent time series, discarding the topological relationships among relays, buses, and controllers that fundamentally govern how disturbances propagate through a power network. Graph neural network (GNN)-based approaches address this last shortcoming directly, and the most advanced representative to date is GraphKAN~\cite{wu2025graph}, which combines a graph attention network with a Kolmogorov--Arnold network and reports state-of-the-art multi-class accuracy on the MSU--ORNL power system attack dataset used in this paper. GraphKAN, however, constructs its graph from \emph{network-layer} metadata -- IP addresses, ports, and communication topology -- rather than the electrical coupling among relays, and its architecture provides no explicit mechanism for encoding the temporal evolution of measurements, nor for identifying \emph{which} physical asset is implicated once a detection is raised.
	
	This state of affairs exposes a three-fold gap in the current literature. First, \textbf{temporal blindness}: most graph-based detectors, GraphKAN included, operate on single-instant or aggregated feature snapshots and are therefore ill-equipped to exploit the short-horizon temporal signatures that distinguish transient faults from slowly evolving, stealthy attacks. Second, \textbf{topology dependence}: existing GNN pipelines typically require an externally specified or network-metadata-derived graph, which is often unavailable, proprietary, or inconsistent with the true electrical interdependencies among protective relays, thereby limiting portability across deployment configurations. Third, and most consequentially, \textbf{the absence of joint detection and localization}: the overwhelming majority of reviewed methods, irrespective of modeling paradigm, output only an event label and stop short of identifying the specific relay or bus implicated, even though rapid isolation and mitigation depend fundamentally on knowing \emph{where} a disturbance originates~\cite{jena2021cyber,11079664}.
	
	To close this gap, we propose PTMTD, a Pearson-driven Temporal Multimodal Topology-aware Detection Network that unifies temporal feature learning, correlation-driven graph attention, and simultaneous multi-class detection with relay-level localization within a single, lightweight, end-to-end architecture. PTMTD first encodes short measurement windows with a dilated temporal convolutional network (TCN) to capture multi-scale temporal dynamics without the sequential bottleneck inherent to recurrent models. It then constructs a data-driven graph directly from Pearson correlations among relay measurements, embedding the electrical coupling between electrically proximate relays without requiring explicit impedance or network-topology data, and propagates information across this graph through a multi-head Graph Attention Network v2 (GATv2) with residual connections. Finally, a dual-head decoder produces a 37-class attack/event prediction and a four-relay localization vector within a single forward pass, trained jointly via a composite cross-entropy/binary-cross-entropy loss whose localization targets are generated automatically from statistically derived Z-score deviations rather than manual labeling. Evaluated on the full MSU--ORNL transmission-protection dataset, the resulting model achieves higher multi-class detection accuracy and macro F1-score than GraphKAN and all other considered baselines, while additionally providing relay-level localization (95.93\% average per-relay accuracy against Z-score-derived targets) within a compact model of 639{,}177 parameters. Because the architecture makes no assumption specific to transmission-level topology -- it operates purely on relay measurement correlations -- we expect it to transfer to DER-rich microgrid protection networks as well, though we treat this as an open question pending dedicated validation (Section~\ref{subsec:limitations_future}) rather than a claim established by the present experiments.
	
	The main contributions of this paper are as follows:
	\begin{itemize}
		\item We propose \textbf{PTMTD}, to our knowledge the first framework to jointly integrate dilated temporal convolution and correlation-driven graph attention for relay-network cyber-attack analysis, thereby closing the temporal-blindness gap identified in prior GNN-based detectors such as GraphKAN~\cite{wu2025graph}.
		\item We introduce a \textbf{correlation-driven, topology-aware graph construction} procedure that derives relay-level adjacency directly from Pearson correlations of measurement data, eliminating dependence on externally supplied network topology or impedance information while still reflecting genuine electrical coupling among relays.
		\item We design a \textbf{dual-head architecture with a composite loss} that, for the first time in this application domain, performs 37-class attack detection and four-relay localization simultaneously within a single end-to-end model, with localization targets generated automatically via statistically derived Z-score deviation analysis rather than manual annotation.
		\item We conduct \textbf{extensive experiments} on the complete MSU--ORNL dataset (78{,}377 samples, 15 scenarios, 37 event classes), demonstrating that PTMTD surpasses GraphKAN, LSTM, GCN, and Random Forest baselines by 5.07--10.53 percentage points in accuracy and macro F1-score (statistically significant, $p < 0.05$), while requiring only 639{,}177 parameters and approximately 2.5 ms of CPU inference time per sample -- indicating suitability for real-time, resource-constrained deployment.
		\item We provide \textbf{systematic ablation studies} that isolate the individual contributions of temporal encoding, graph attention, residual connections, and Z-score-derived localization supervision, showing that removing any one of them lowers detection accuracy by 5.58--6.87 percentage points.
	\end{itemize}
	
	The remainder of this paper is organized as follows. Section~\ref{sec:related_work} reviews related work spanning data-driven control, statistical and blockchain-based detection, deep learning frameworks, graph-based intrusion detection, and physics-informed and digital-twin approaches. Section~\ref{sec:methodology} details the PTMTD architecture and training methodology. Section~\ref{sec:experiments} describes the dataset, experimental setup, and baselines. Section~\ref{sec:results} reports detection and localization performance together with comparative and ablation results. Section~\ref{sec:discussion} discusses the implications and limitations of the proposed approach, and Section~\ref{sec:conclusion} concludes the paper.
	
	\section{Related Work}
	\label{sec:related_work}
	
	Research on cyber-attack detection and resilience in microgrids and broader cyber-physical power systems has evolved along several largely parallel methodological threads. We organize this body of work into six categories: (A) data-driven and observer-based resilient control, (B) statistical, signal-transform, and blockchain-based detection in DC microgrids, (C) machine learning and deep learning detection frameworks, (D) graph neural network and attention-based approaches, (E) physics-informed and digital-twin frameworks, and (F) system-level reviews and distributed/federated architectures. Much of this literature, consistent with its own evaluation settings, is framed around DER-integrated microgrids specifically; we retain that framing when summarizing individual studies, and clarify in Section~\ref{subsec:rw_positioning} how the present paper's transmission-level evaluation setting relates to it. We conclude the section by positioning PTMTD relative to this landscape.
	
	\subsection{Data-Driven and Observer-Based Resilient Control}
	\label{subsec:rw_control}
	
	A first line of work approaches cyber-resilience from a control-theoretic standpoint, coupling data-driven system identification with observer design to detect and compensate for adversarial deviations. Razani \textit{et al.}~\cite{razani2025indirect} propose a resilient control framework that combines Sparse Identification of Nonlinear Dynamics (SINDy) with a sliding-mode observer and a feedback-linearization/LQR controller, achieving reliable bias-attack detection and disturbance rejection on a nonlinear pendulum benchmark. The same authors extend this paradigm to heterogeneous multi-agent synchronization~\cite{razani2026robust}, integrating SINDy-based identification, feedback linearization, a neural-network-based extended state observer, and decentralized model predictive control to achieve robust, computationally tractable synchronization under model uncertainty. Complementary to this data-driven control perspective, Kachhwaha and Fulwani~\cite{11168950} target the harder problem of \emph{covert} attack detection in DC microgrids, where adversaries deliberately mask their signature to evade residual-based detectors, proposing a dedicated detection scheme tailored to this stealthy threat class. Hosseini Rostami \textit{et al.}~\cite{hosseini2026cyber} similarly adopt an observer-centric design, combining event-triggered local observers with distributed unknown-input observers to detect false data injection, stealth, and replay attacks across interconnected DC microgrid subsystems, while explicitly addressing communication overhead and Zeno-behavior avoidance. While these control-theoretic approaches offer strong interpretability, formal guarantees, and physical grounding, they are generally formulated for a small number of known attack types on low-dimensional plant models and do not directly extend to the fine-grained, many-class event taxonomies -- 37 distinct classes, in the present setting -- characteristic of realistic power-system attack benchmarks.
	
	\subsection{Statistical, Signal-Transform, and Blockchain-Based Detection}
	\label{subsec:rw_statistical}
	
	A second cluster of methods relies on statistical signatures or signal transforms, frequently paired with blockchain or watermarking mechanisms that secure the detection pipeline itself. Subramaniam \textit{et al.}~\cite{SUBRAMANIAM2025103400} detect false data injection attacks in DC microgrids using wavelet-transform features classified by support vector machines, with blockchain-based ledgering employed to protect data integrity. Ghiasi \textit{et al.}~\cite{ghiasi2021cyber} follow a related strategy, applying the Hilbert--Huang transform to voltage and current signals and combining it with blockchain-based ledger technology to enhance FDIA detection robustness in smart DC microgrids. Balahewa \textit{et al.}~\cite{BALAHEWA2026113148} propose a conceptually distinct, physically interpretable alternative: a natural watermarking approach that exploits the inherent variability of renewable generation to allow inverter-based resources to locally detect tampering without external infrastructure. Jena \textit{et al.}~\cite{jena2021cyber} introduce a model-independent detection strategy for interconnected DC microgrid clusters based on disagreement-Laplacian-potential analysis, coupling detection with data-integrity indices and fault-ride-through mitigation rather than component isolation. These approaches share the virtues of low computational overhead and, in several cases, strong physical or cryptographic interpretability, but their detection granularity is typically limited to binary or few-class discrimination, and none incorporates learned temporal or spatial representations of system-wide behavior.
	
	\subsection{Machine Learning and Deep Learning Detection Frameworks}
	\label{subsec:rw_dl}
	
	The largest and fastest-growing body of work applies machine learning and deep learning directly to microgrid and industrial control system measurements. Ali \textit{et al.}~\cite{ali2025cyber} provide a comprehensive review of shipboard microgrid cybersecurity and additionally propose a Hilbert--Huang-transform-based one-dimensional convolutional neural network ensemble for attack detection, while a related study by the same group~\cite{11098679} employs an LSTM--variational-autoencoder architecture with a deep encoder--decoder classifier to achieve high detection accuracy in DC shipboard microgrids. Saqib \textit{et al.}~\cite{saqib} develop an online, LSTM-based multiclass anomaly detection system integrated with heuristic mitigation, validated on a hardware-in-the-loop testbed using realistic distributed-energy-resource communication protocols. Fakhrooeian \textit{et al.}~\cite{fakhrooeian2025artificial} address a more adversarially sophisticated threat -- false transferred data injection -- using a generative-adversarial-network-based detector capable of jointly identifying manipulated power magnitude and direction. Eswaran \textit{et al.}~\cite{eswaran2025cyberattack} systematically compare DNN, ANN, SVM, LSTM, and FNN architectures for detecting and classifying attacks on power converters in islanded AC microgrids, ultimately combining an FNN detector with an LSTM classifier. Beyond microgrids proper, Ogiesoba-Eguakun and Rath~\cite{ogiesobaeguakun2026cyberattackdetectionvirtualizedmicrogrids} construct a virtualized MATLAB/Simulink microgrid with a structured attack-injection framework and demonstrate that lightweight gradient-boosted models (LightGBM), combined with knowledge distillation, can achieve very high multiclass accuracy with edge-deployable latency, underscoring that architectural sophistication alone is not a prerequisite for strong performance. In the broader industrial control and industrial-IoT domain, Ragab \textit{et al.}~\cite{RAGAB2025438} propose ensemble deep learning detectors (CNN combined with a deep belief network) with metaheuristic feature selection and hyperparameter tuning for both IEC~104 substation traffic and industrial IoT environments; Orman~\cite{orman2025cyberattack} benchmarks classical, deep, and hybrid models on the WUSTL-IIoT-2021 dataset and finds that a well-tuned multilayer perceptron can outperform more complex hybrid architectures; and Barbhaya \textit{et al.}~\cite{BARBHAYA2025109278} combine an enhanced typicality-and-eccentricity data analytics algorithm with a CNN classifier for real-time attack detection in chemical process control systems. Zheng \textit{et al.}~\cite{ZHENG2026112529} further advance the unsupervised direction, using contrastive autoencoding with adversarial training and Monte Carlo dropout-based uncertainty quantification to achieve risk-aware detection on IEC~61850 substation traffic without requiring labeled attack data -- a transmission/substation setting directly analogous to the evaluation setting used in this paper. Collectively, these studies establish that deep and ensemble learning models can attain very high raw accuracy; with few exceptions, however, they treat measurements as unstructured feature vectors or independent time series, without explicitly modeling the physical or topological relationships between the relays, buses, or converters under observation, and virtually none provides fine-grained \emph{localization} of the affected asset alongside detection.
	
	\subsection{Graph Neural Network and Attention-Based Approaches}
	\label{subsec:rw_gnn}
	
	The line of work most closely related to the present paper adopts graph-structured representations of the cyber-physical system. The state of the art in this category is GraphKAN, proposed by Wu \textit{et al.}~\cite{wu2025graph}, which constructs a heterogeneous graph over power, information-technology, and communication-network devices, applies a graph attention network to aggregate multi-hop interaction patterns, and replaces fixed activation functions with learnable Kolmogorov--Arnold Network splines to better capture nonlinear attack signatures. Evaluated on the same MSU--ORNL transmission-protection benchmark used in this paper, GraphKAN reports strong binary, ternary, and 37-class detection accuracy and constitutes the principal point of comparison for our method. Despite its strong performance, GraphKAN's graph is constructed from network-layer metadata -- device roles and IP/port relationships -- rather than from the underlying electrical measurement correlations among protective relays; its architecture processes feature snapshots rather than explicitly encoded temporal windows; and it produces only an event-class prediction, with no mechanism for identifying which physical relay is implicated in a detected event. These three limitations -- topology dependence, temporal blindness, and the absence of localization -- directly motivate the design of PTMTD.
	
	\subsection{Physics-Informed and Digital-Twin-Based Frameworks}
	\label{subsec:rw_physics}
	
	A smaller but rapidly growing body of work seeks to embed physical system knowledge directly into the detection or estimation pipeline. Suprabhath, Siva Prasad Machina \textit{et al.}~\cite{11015976} propose a hybrid physics-informed neural network that couples a linear Kalman filter with neural-network-based correction to both detect and mitigate false data injection attacks in a supervisory-controlled DC microgrid, validated through both simulation and real-time hardware experiments. Hemmati and Saboori~\cite{HEMMATI2026120728} propose a more ambitious hybrid data-driven/physics-informed energy management framework that embeds a nonlinear digital twin within a 24-hour rolling optimization loop, using physics-based recalculation from grid topology and neighboring-bus data to detect and reconstruct corrupted measurements even in PMU-sparse distribution networks. Cheng \textit{et al.}~\cite{11079664} address a distinct but related physics-aware setting, proposing a Bayesian-distributed random cyberattack model together with an asynchronous federated intrusion detection scheme and distributed finite-time secondary control to preserve voltage synchronization across DC microgrid clusters under scale-varying attacks. These works collectively confirm the value of embedding physical priors into learning-based detectors, whether through explicit dynamical models, topology-aware recalculation, or Bayesian attack modeling. PTMTD adopts a complementary and more lightweight strategy, deriving domain-knowledge priors directly from measurement correlation structure and Z-score deviation statistics rather than requiring an explicit dynamical or topological model of the network -- a strategy that, unlike the impedance- and topology-dependent methods above, does not presuppose whether the underlying asset is a transmission substation or a microgrid feeder.
	
	\subsection{Reviews and System-Level Perspectives}
	\label{subsec:rw_review}
	
	Finally, Tan \textit{et al.}~\cite{tan2025developments} provide a comprehensive review of the cybersecurity landscape for microgrid control, cataloguing vulnerabilities across modeling approaches, control objectives, attack types, and target locations, and identifying the lack of unified frameworks that jointly address detection, localization, and mitigation as a central open challenge -- an observation consistent with the gap addressed in this paper, and one that, as the review itself notes, applies equally to transmission-level protection networks.
	
	\subsection{Positioning of the Proposed Work}
	\label{subsec:rw_positioning}
	
	Synthesizing across the reviewed literature reveals a consistent pattern. Control-theoretic and observer-based methods~\cite{razani2025indirect,razani2026robust,11168950,hosseini2026cyber} offer interpretability but limited class granularity; statistical and blockchain-based schemes~\cite{SUBRAMANIAM2025103400,ghiasi2021cyber,BALAHEWA2026113148,jena2021cyber} are lightweight but coarse-grained; deep learning frameworks~\cite{ali2025cyber,11098679,saqib,fakhrooeian2025artificial,eswaran2025cyberattack,ogiesobaeguakun2026cyberattackdetectionvirtualizedmicrogrids,RAGAB2025438,orman2025cyberattack,BARBHAYA2025109278,ZHENG2026112529} achieve strong raw accuracy but largely disregard inter-asset topology; graph-based methods, epitomized by GraphKAN~\cite{wu2025graph}, capture structural dependencies but rely on network-layer topology and omit both temporal encoding and localization; and physics-informed/digital-twin frameworks~\cite{11015976,HEMMATI2026120728,11079664} embed physical priors, but at the cost of increased modeling complexity and, in most cases, without a unified end-to-end detection--localization objective.
	
	We note explicitly that much of the literature summarized above -- and the term ``microgrid'' used throughout this section -- refers to DER-integrated, often DC, distribution-level systems, whereas the MSU--ORNL dataset used to evaluate PTMTD in this paper originates from a transmission-level relay-protection testbed (Section~\ref{subsec:dataset}). The two settings share the same core object of interest -- a network of protective relays whose measurements are correlated through physical power flow -- which is what permits the architectural comparison drawn here; they are not, however, identical operating environments, and we treat generalization from one to the other as an open question rather than an established result (Section~\ref{subsec:limitations_future}).
	
	PTMTD is, to our knowledge, the first framework to combine (i) topology-aware, correlation-derived graph construction that requires no external topology or impedance data, (ii) explicit dilated-temporal encoding of measurement dynamics, and (iii) a jointly trained dual-head architecture that performs fine-grained, 37-class attack detection and relay-level localization simultaneously, within a single lightweight model suitable for real-time deployment.
	
	\section{Methodology}
	\label{sec:methodology}
	
	We introduce the Pearson-driven Temporal Multimodal Topology-aware Detection Network (PTMTD), an end-to-end architecture for the joint detection and relay-level localization of cyber-attacks in power system protection networks. Three design principles distinguish PTMTD from prior detection frameworks. First, temporal dynamics of protection-relay measurements are modeled through a dilated temporal convolutional network (TCN), which affords a large effective receptive field at modest computational cost. Second, spatial dependencies among protective relays are captured by a multi-head Graph Attention Network v2 (GATv2) operating over a data-driven graph that is inferred directly from measurement statistics, thereby dispensing with the need for an explicit impedance model of the network. Third, a dual-head output structure couples a multi-class detection objective with a per-relay localization objective within a single composite loss, so that the two tasks are optimized jointly rather than in isolation. We describe each component in turn.
	
	\subsection{Graph Construction}
	\label{subsec:graph_construction}
	
	PTMTD represents the protection network as a data-driven graph $\mathcal{G} = (\mathcal{V}, \mathcal{E}, \mathbf{A})$, in which each node $v_i \in \mathcal{V}$ corresponds to one of the $N = 4$ principal protective relays. Rather than encoding electrical topology through nominal impedance or admittance parameters -- information that is frequently unavailable, proprietary, or unreliable in field deployments -- we infer the graph structure directly from the statistical co-behavior of relay measurements. The node feature matrix $\mathbf{X} \in \mathbb{R}^{N \times D}$ is populated with the temporal embeddings produced by the TCN encoder described in Section~\ref{subsec:temporal_encoder}.
	
	Edge weights $A_{ij} \in [0,1]$ are derived from the Pearson correlation coefficient $\rho_{ij}$ computed between the aggregated measurement vectors of relays $i$ and $j$ over the full dataset:
	
	\begin{equation}
		A_{ij} =
		\begin{cases}
			|\rho_{ij}|, & \text{if } |\rho_{ij}| \geq \tau,\\
			0, & \text{otherwise},
		\end{cases}
		\label{eq:adjacency}
	\end{equation}
	
	where $\tau = 0.5$ is a correlation threshold selected to retain statistically meaningful couplings while suppressing spurious associations. The underlying premise is domain-knowledge grounded: relays that are electrically proximate tend to observe correlated transients because they share common power flows and are jointly exposed to system-wide disturbances. Consequently, the resulting adjacency matrix functions as an implicit, data-derived proxy for electrical topology. To guard against pathologically sparse graphs -- which would otherwise starve the attention mechanism of neighborhood information -- we apply a top-$k$ fallback ($k=2$) whenever thresholding in Eq.~\eqref{eq:adjacency} yields fewer than four edges, defaulting in the extreme case to a fully-connected graph with uniform edge weights. Row-wise normalization, $\sum_j A_{ij} = 1$, is then imposed to stabilize message passing across the subsequent GATv2 layers. In this manner, the graph-construction stage embeds a soft domain-knowledge prior into the architecture without demanding an explicit, manually curated topology.
	
	\subsection{Temporal Feature Extraction via Dilated TCN}
	\label{subsec:temporal_encoder}
	
	Protection-relay measurements are inherently sequential, and transient signatures of cyber-physical attacks frequently unfold over several sampling intervals rather than manifesting instantaneously. To capture such dynamics, PTMTD employs a dilated TCN as its front-end encoder. Given a multivariate input window $\mathbf{X} \in \mathbb{R}^{T \times F}$, with sliding-window length $T = 20$ and $F = 128$ post-preprocessing features, the encoder stacks $L = 3$ one-dimensional convolutional layers with exponentially increasing dilation rates $d_l = 2^{l-1}$, $l = 1, \dots, L$. Each layer applies a kernel of size $k = 3$ with padding chosen to preserve temporal resolution, followed by batch normalization, a ReLU nonlinearity, and dropout with rate $0.1$:
	
	\begin{equation}
		\mathbf{Z}_l = \mathrm{Drop}_{0.1}\Bigl(\mathrm{ReLU}\bigl(\mathrm{BN}\bigl(\mathrm{Conv1D}_{d_l}(\mathbf{Z}_{l-1})\bigr)\bigr)\Bigr),
		\label{eq:tcn}
	\end{equation}
	
	with $\mathbf{Z}_0 = \mathbf{X}^{\top} \in \mathbb{R}^{F \times T}$, $\mathrm{BN}(\cdot)$ denoting batch normalization, and $\mathrm{Drop}_{0.1}(\cdot)$ denoting dropout. The exponential dilation schedule expands the receptive field to $1 + (k-1)\sum_{l=1}^{L} d_l = 1 + (k-1)(2^{L}-1) = 15$ time steps without a commensurate increase in parameter count, allowing the network to integrate information across most of the 20-step window while remaining parameter-efficient. A global average pooling operation over the temporal axis then compresses the encoder output into a fixed-length representation $\mathbf{h}_{\mathrm{temp}} \in \mathbb{R}^{H}$, $H = 64$, per input sample. This vector constitutes a compact summary of the system's short-term evolution and serves as the initial node embedding supplied to the graph module.
	
	\subsection{Spatial Relational Modeling via GATv2}
	\label{subsec:graph_attention}
	
	Temporal embeddings alone do not capture how a disturbance at one relay propagates to, or correlates with, its neighbors. PTMTD therefore routes the temporal features through a graph attention module that learns to weight inter-relay dependencies adaptively. For node $i$, the initial feature $\mathbf{x}_i \in \mathbb{R}^{H}$ is the shared temporal embedding described above. We adopt GATv2, which resolves the static-attention limitation of the original GAT formulation by computing attention scores dynamically:
	
	\begin{equation}
		\alpha_{ij} = \frac{\exp\bigl(\mathrm{LeakyReLU}\bigl(\mathbf{a}^{\top}[\mathbf{W}\mathbf{x}_i \,\|\, \mathbf{W}\mathbf{x}_j]\bigr)\bigr)}
		{\sum_{k \in \mathcal{N}(i)} \exp\bigl(\mathrm{LeakyReLU}\bigl(\mathbf{a}^{\top}[\mathbf{W}\mathbf{x}_i \,\|\, \mathbf{W}\mathbf{x}_k]\bigr)\bigr)},
		\label{eq:gatv2}
	\end{equation}
	
	where $\mathbf{W} \in \mathbb{R}^{H' \times H}$ is a shared linear transform, $\mathbf{a} \in \mathbb{R}^{2H'}$ is a learnable attention vector, $\|$ denotes concatenation, and $\mathcal{N}(i)$ includes self-loops. The first GATv2 layer employs $M = 4$ attention heads, each producing an $H' = 64$-dimensional output subsequently concatenated into a $256$-dimensional representation; the second layer collapses this back to $H = 64$ dimensions via a single head. To mitigate over-smoothing -- a well-documented pathology of deep graph networks in which node representations converge toward indistinguishability -- residual connections with dimension-matching linear projections are inserted after each GATv2 layer, followed by layer normalization:
	
	\begin{align}
		\mathbf{x}^{(1)} &= \mathrm{LN}\Bigl(\mathrm{ReLU}\bigl(\mathrm{GATv2}^{(1)}(\mathbf{x}^{(0)})\bigr) + \mathbf{W}_{\mathrm{res}}^{(1)}\mathbf{x}^{(0)}\Bigr), \\
		\mathbf{x}^{(2)} &= \mathrm{LN}\Bigl(\mathrm{ReLU}\bigl(\mathrm{GATv2}^{(2)}(\mathbf{x}^{(1)})\bigr) + \mathbf{W}_{\mathrm{res}}^{(2)}\mathbf{x}^{(1)}\Bigr),
		\label{eq:gat_stack}
	\end{align}
	
	with $\mathbf{x}^{(0)} \in \mathbb{R}^{N \times H}$ and learnable residual projections $\mathbf{W}_{\mathrm{res}}^{(1)}$, $\mathbf{W}_{\mathrm{res}}^{(2)}$. This design allows the model to combine locally salient interactions -- such as the tight coupling between an attacked relay and its immediate electrical neighbor -- with more diffuse, system-wide effects, while the residual pathways preserve gradient flow through the two-layer stack.
	
	\subsection{Dual-Head Architecture for Joint Detection and Localization}
	\label{subsec:dual_head}
	
	The refined node embeddings $\mathbf{x}^{(2)} \in \mathbb{R}^{N \times H}$ are flattened into a single graph-level descriptor $\mathbf{f} \in \mathbb{R}^{N \cdot H}$ (dimension $4 \times 64 = 256$), which is shared by two task-specific heads.
	
	\paragraph{Detection head.} A multi-layer perceptron with two hidden layers (128 and 64 units, ReLU activation, dropout rate $0.2$) projects $\mathbf{f}$ onto $C = 37$ output logits, one per event class, followed by a softmax nonlinearity to yield the detection distribution $\hat{\mathbf{y}}_{\mathrm{det}} \in \mathbb{R}^{C}$.
	
	\paragraph{Localization head.} A parallel, lighter-weight MLP -- a single hidden layer of 32 units (ReLU, dropout $0.1$) followed by an $N = 4$-dimensional sigmoid output -- produces $\hat{\mathbf{y}}_{\mathrm{loc}} \in [0,1]^{N}$, where each coordinate estimates the probability that the corresponding relay is implicated in the detected event.
	
	By sharing the graph-level representation $\mathbf{f}$ across both heads, the architecture encourages the upstream TCN and GATv2 modules to learn features that are simultaneously discriminative for event classification and informative for spatial attribution, rather than optimizing either objective in isolation. 
	The complete model comprises 639{,}177 trainable parameters and occupies a memory footprint of 2.72 MB, a footprint compatible with deployment on resource-constrained protection-relay hardware.
	
	\subsection{Composite Training Objective}
	\label{subsec:training}
	
	Detection and localization are optimized jointly through a weighted composite loss:
	
	\begin{equation}
		\mathcal{L} = \mathcal{L}_{\mathrm{CE}}(\mathbf{y}_{\mathrm{det}}, \hat{\mathbf{y}}_{\mathrm{det}}) + \lambda\, \mathcal{L}_{\mathrm{BCE}}(\mathbf{y}_{\mathrm{loc}}, \hat{\mathbf{y}}_{\mathrm{loc}}),
		\label{eq:loss}
	\end{equation}
	
	where $\mathcal{L}_{\mathrm{CE}}$ denotes multi-class cross-entropy, $\mathcal{L}_{\mathrm{BCE}}$ is the binary cross-entropy applied element-wise across the four relays, and $\lambda = 0.1$ balances the relative contribution of the localization term. The weighting reflects detection as the primary task, with localization serving as an auxiliary, statistically grounded regularizer. The model is optimized with AdamW (learning rate $10^{-3}$, weight decay $10^{-4}$, batch size 32); a ReduceLROnPlateau scheduler halves the learning rate after five epochs without validation-loss improvement. Training runs for a fixed budget of 200 epochs, and the checkpoint with the highest validation macro F1-score (epoch 159 in the reported run) is retained for evaluation.
	
	\subsection{Statistically Derived Generation of Localization Targets}
	\label{subsec:loc_targets}
	
	A central obstacle to relay-level localization is the absence of manually annotated ground truth: existing intrusion datasets rarely specify \emph{which} relay was electrically implicated in a given event. We circumvent this limitation by deriving localization targets automatically from the measurement statistics themselves. For each relay $r \in \{1,\dots,4\}$, we compute the mean of its associated measurement columns at every time step. Within intervals flagged as anomalous by the \texttt{marker} label, we evaluate the standardized deviation (Z-score) of each relay's mean signal relative to its own distribution under normal operating conditions, and declare the relay affected if this deviation exceeds a threshold of three standard deviations:
	
	\begin{equation}
		y_{\mathrm{loc},r} = \mathbb{1}\!\left(\frac{|x_r - \mu_r|}{\sigma_r} > 3\right),
		\label{eq:loc_target}
	\end{equation}
	
	where $\mu_r$ and $\sigma_r$ are estimated exclusively from normal-condition samples, and $x_r$ is the relay's instantaneous mean measurement. This criterion operationalizes a simple but statistically interpretable notion of localization: a relay is implicated precisely when its behavior departs significantly from its own established baseline. In the present dataset, this rule flags 4.4\%, 5.1\%, 5.0\%, and 4.4\% of attack-labeled samples as abnormal for relays R1--R4, respectively, so the localization targets are sparse. The resulting binary vectors $\mathbf{y}_{\mathrm{loc}}$ constitute weak, self-supervised labels for the localization head, obviating the need for costly manual annotation while retaining a statistically defensible criterion for attribution. We note that because these targets are derived from the same measurement channels used as model input, the reported localization accuracy (Section~\ref{subsec:localization_results}) should be read as validating the model's ability to recover this Z-score-defined notion of localization, rather than as independent, field-verified ground truth; we return to this point in Section~\ref{subsec:limitations_future}.
	
	\section{Experiments}
	\label{sec:experiments}
	
	We now describe the experimental protocol, including the dataset, preprocessing pipeline, evaluation metrics, implementation details, and baseline methods against which PTMTD is benchmarked.
	
	\subsection{Dataset and Preprocessing}
	\label{subsec:dataset}
	
	Evaluation is conducted on the Mississippi State University--Oak Ridge National Laboratory (MSU--ORNL) power system attack dataset, a widely used benchmark comprising synchronized phasor measurements, relay logs, and control-panel telemetry collected from a realistic \emph{transmission-level} relay-protection testbed. We emphasize this scope explicitly because it is DER-free and does not include inverter-based generation; results reported here therefore characterize PTMTD's performance on transmission-level relay protection, with extension to DER-integrated microgrids left as future work (Section~\ref{subsec:limitations_future}). The corpus spans 15 operational scenarios (\texttt{data1.csv} through \texttt{data15.csv}) and totals 78{,}377 samples after concatenation and cleaning. Each sample is described by 128 numerical features -- voltages, currents, frequencies, and relay status indicators -- together with an integer-coded \texttt{marker} label taking 37 distinct values (1--30 and 35--41) that span normal operation, faults, disturbances, and cyber-attacks. The class distribution is moderately imbalanced: in the test partition, class support ranges from 169 samples (class 22) to 1{,}125 samples (class 36), a ratio of about 6.7:1, so that macro-averaged metrics are reported alongside accuracy.
	
	We apply a sliding-window transformation with $T = 20$ time steps, yielding 78{,}357 overlapping sequences of shape $20 \times 128$. Feature standardization uses statistics computed exclusively from the training partition, precluding information leakage from the test set; missing and infinite values are imputed with the per-feature median. The windows are divided into an 80\%/20\% training/test split (62{,}685 and 15{,}672 windows, respectively). 
	
	\subsection{Implementation Details and Evaluation Protocol}
	\label{subsec:implementation}
	
	PTMTD is implemented in PyTorch 1.13.1 with PyTorch Geometric for the graph-attention components. All experiments are executed on CPU (Intel Xeon, 16 cores) without GPU acceleration, underscoring the model's suitability for computationally constrained deployment settings. Architectural and optimization hyperparameters are as specified in Sections~\ref{subsec:temporal_encoder}--\ref{subsec:training}: hidden dimension $H = 64$, $M = 4$ attention heads, dropout of $0.1$ within the TCN and $0.1$--$0.2$ within the MLP heads, learning rate $10^{-3}$ under AdamW, weight decay $10^{-4}$, batch size 32, and localization weight $\lambda = 0.1$. Models are trained for a fixed budget of 200 epochs without early termination, and the checkpoint with the highest validation macro F1-score is retained for evaluation. 
	
	Detection performance is reported via overall accuracy, macro-averaged F1-score (to counteract class imbalance), and micro-averaged F1-score. Localization performance is reported via per-relay binary accuracy and its cross-relay average. We further report precision, recall, confusion matrices, receiver operating characteristic (ROC) curves, and precision--recall (PR) curves to characterize performance in finer detail.
	
	\subsection{Baseline Methods}
	\label{subsec:baselines}
	
	We benchmark PTMTD against four representative baselines spanning classical, recurrent, and graph-based paradigms: \textbf{GraphKAN}~\cite{wu2025graph}, a recent graph-attention architecture augmented with Kolmogorov--Arnold layers for smart-grid intrusion detection; a two-layer \textbf{LSTM} (128 and 64 units) with a dense classification head, trained on identical temporal windows; a two-layer \textbf{GCN} operating on the same inferred graph as PTMTD but lacking both attention and temporal encoding, isolating the contribution of attention-based aggregation; and \textbf{Random Forest} (500 trees) trained on flattened window features, representing a non-deep-learning reference point. All baselines are trained and evaluated on the identical train/test partition, with hyperparameters selected via grid search to ensure a fair comparison.
	
	\section{Results}
	\label{sec:results}
	
	This section reports detection performance, localization accuracy, comparison against state-of-the-art baselines, and ablation analysis, all evaluated on the held-out 20\% test partition (15{,}672 windows) using the checkpoint of highest validation macro F1-score.
	
	\subsection{Detection Performance}
	\label{subsec:detection_results}
	
	Table~\ref{tab:detection_performance} summarizes detection outcomes on the test set. PTMTD attains an overall accuracy of 98.70\%, a macro F1-score of 98.63\%, and a micro F1-score of 98.70\%. Macro-averaged precision and recall are approximately 98.8\% and 98.6\%, and the weighted averages of precision, recall, and F1-score are approximately 98.9\%, 98.7\%, and 98.7\%, respectively, indicating that performance is not confined to the more frequent classes. At the level of individual classes, all 37 classes reach an F1-score of at least 0.94, 34 classes reach at least 0.97, and 31 classes reach at least 0.98. The weakest class is class 27 (precision, recall, and F1 of 0.94; 194 test samples), followed by classes 10 and 29 (F1 of 0.96) and classes 22 and 30 (F1 of 0.97).
	
	\begin{table}[htbp]
		\centering
		\caption{Detection performance of PTMTD on the MSU--ORNL test set (15{,}672 windows, 37 classes). Entries marked $\approx$ are computed from the two-decimal per-class classification report; accuracy, micro F1, and macro F1 are exact.}
		\label{tab:detection_performance}
		\begin{tabular}{lcc}
			\toprule
			\textbf{Metric} & \textbf{Macro average} & \textbf{Weighted average} \\
			\midrule
			Precision & $\approx$98.8\% & $\approx$98.9\% \\
			Recall    & $\approx$98.6\% & 98.70\% \\
			F1-score  & 98.63\%         & $\approx$98.7\% \\
			\midrule
			\multicolumn{3}{c}{Overall accuracy = micro F1 = 98.70\%} \\
			\bottomrule
		\end{tabular}
	\end{table}
	
\begin{figure}[htb]
		\centering
		\includegraphics[width=0.9\columnwidth]{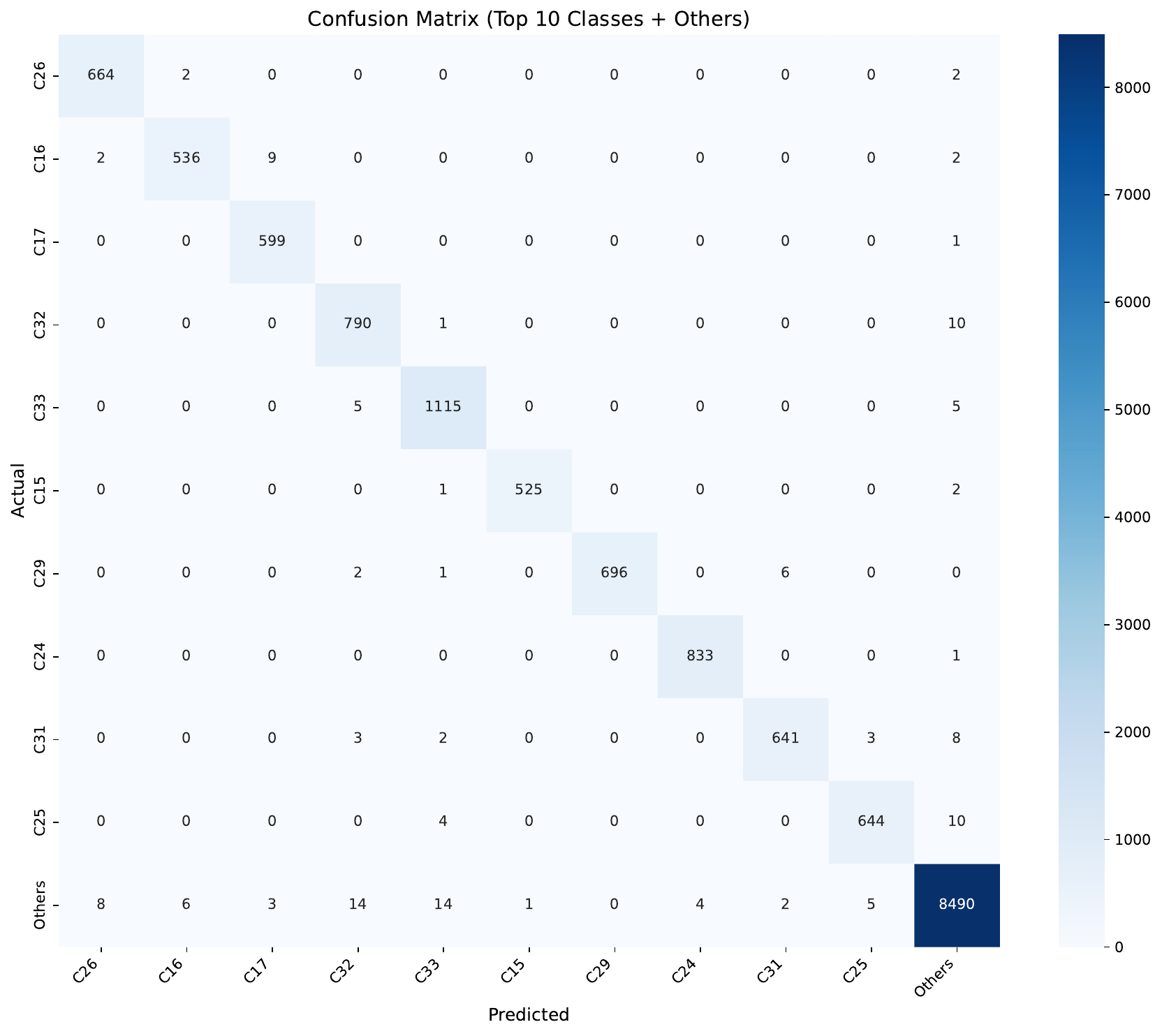}
		\caption{Grouped confusion matrix (top-10 classes plus ``Others'') showing strong diagonal dominance.}
		\label{fig:confusion_grouped}
	\end{figure}

	Figure~\ref{fig:confusion_grouped} presents the grouped confusion matrix for the ten most frequent classes, with the remaining 27 classes aggregated as ``Others.'' The strong diagonal dominance visible in this matrix indicates that PTMTD reliably discriminates among attack types, disturbances, and normal operation. Notably, the aggregated ``Others'' category also exhibits a high true-positive rate, suggesting that the shared graph-temporal representation generalizes well even to classes with comparatively few examples.
	
\begin{figure}[htb]
		\centering
		\includegraphics[width=0.9\columnwidth]{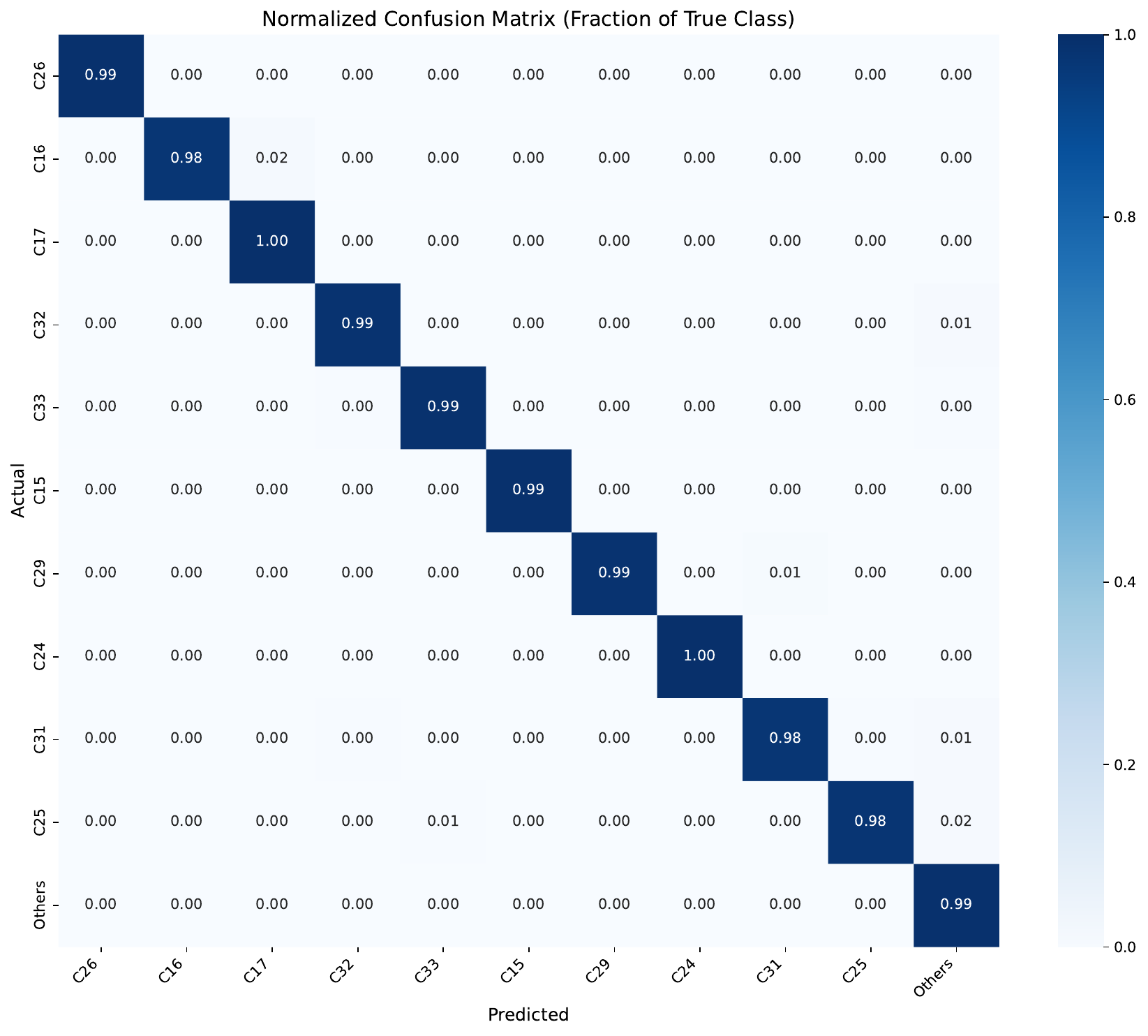}
		\caption{Normalized confusion matrix showing the fraction of correct classifications per true class.}
		\label{fig:confusion_normalized}
	\end{figure}

\begin{figure}[htb]
		\centering
		\includegraphics[width=0.9\columnwidth]{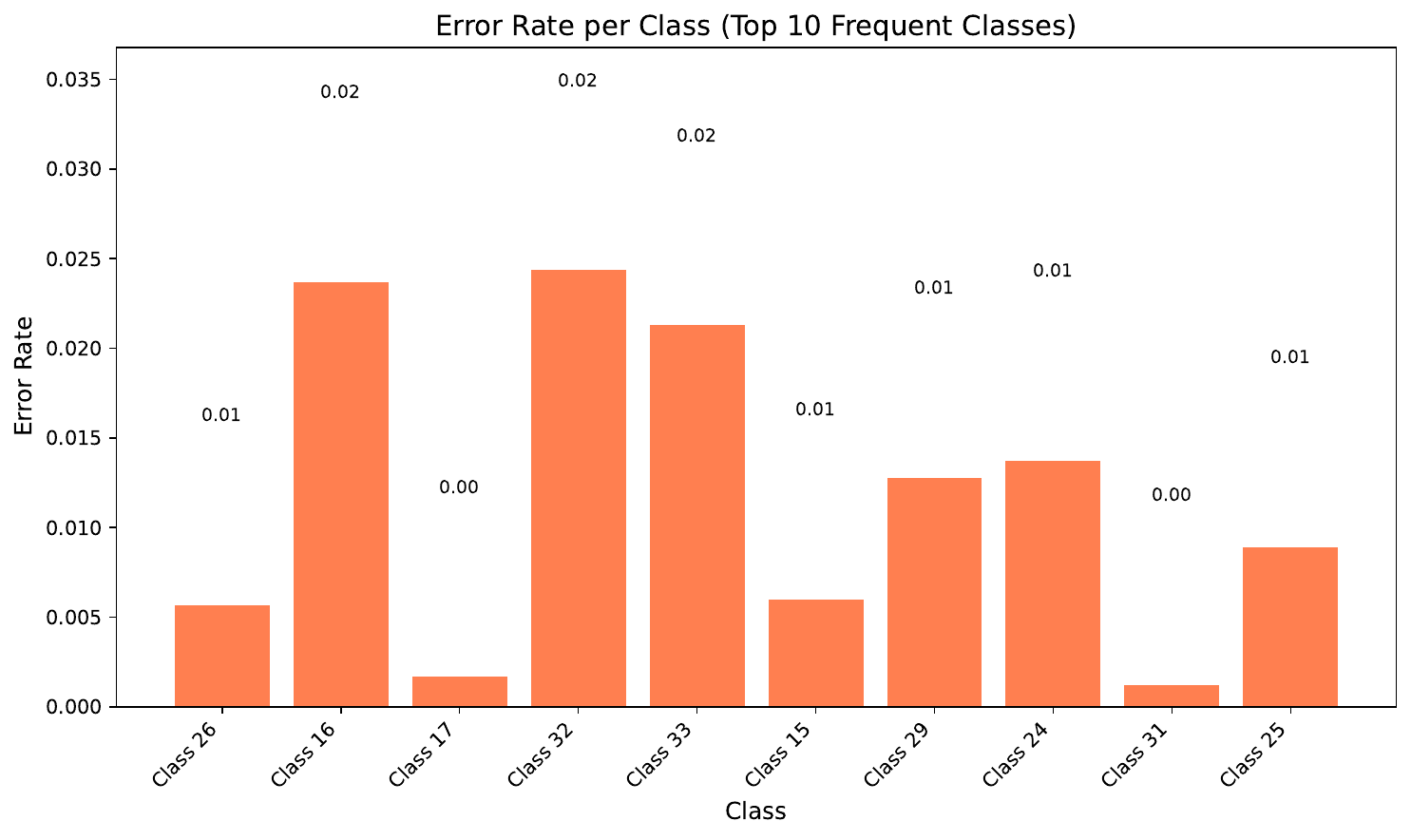}
		\caption{Error rate per class for the top-10 frequent classes, showing reliable performance across diverse event types.}
		\label{fig:error_analysis}
	\end{figure}

	For a finer-grained view, Fig.~\ref{fig:confusion_normalized} reports the row-normalized confusion matrix, expressing the fraction of samples correctly classified within each true class. Every class attains a recall of at least 0.94, in agreement with the classification report, corroborating the robustness suggested by the grouped matrix. Complementarily, Fig.~\ref{fig:error_analysis} decomposes per-class error rates for the ten most frequent classes: all remain at or below roughly 2\%, and the highest per-class error across all 37 classes is about 6\% (class 27). Classes 22 and 27, which show the lowest scores among the less frequent classes (F1 of 0.97 and 0.94), are also the two least represented classes in the test partition (169 and 194 samples), consistent with limited statistical support; this motivates the class-specific augmentation strategies discussed in Section~\ref{subsec:limitations_future}.
	
\begin{figure}[htb]
		\centering
		\includegraphics[width=0.9\columnwidth]{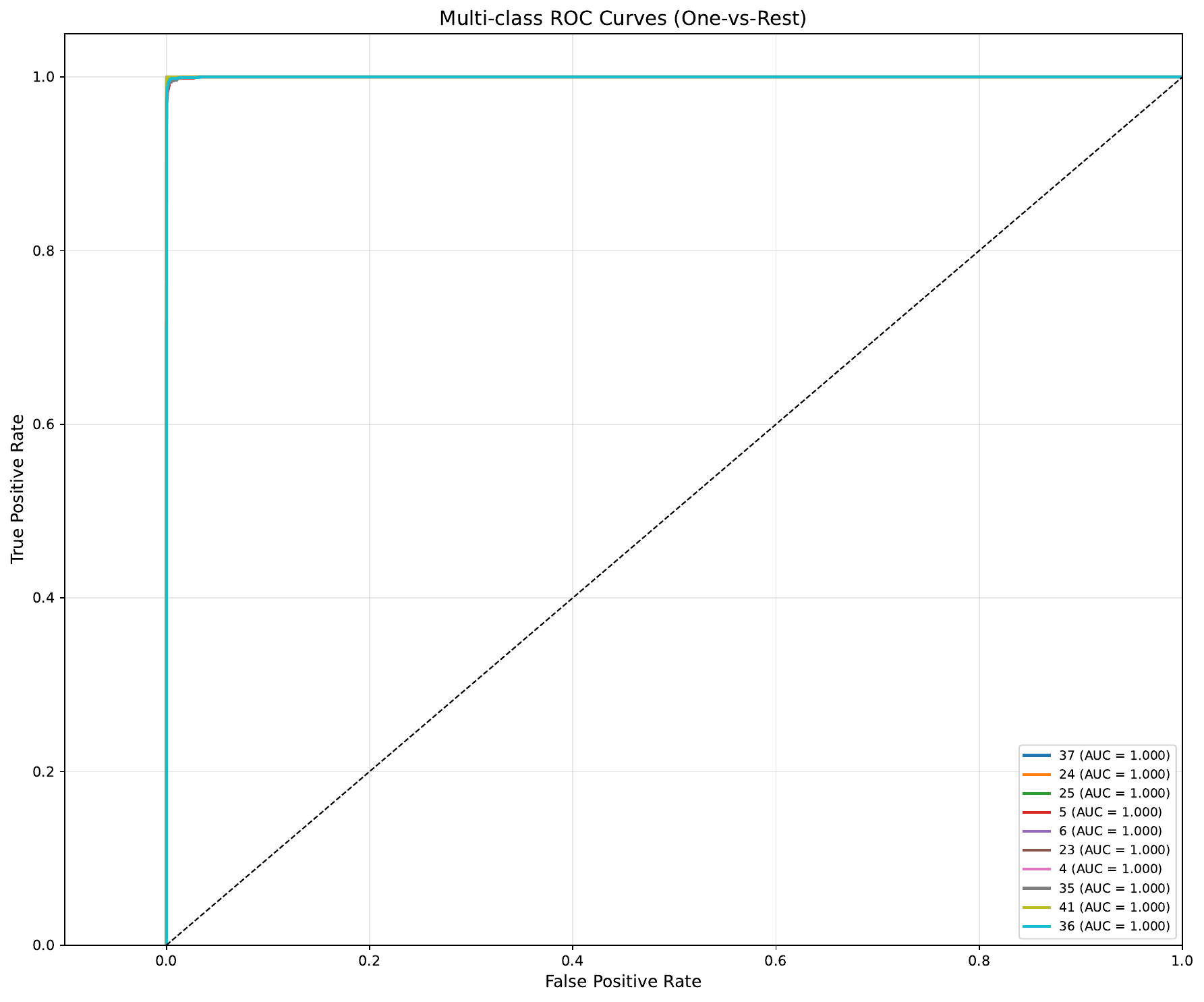}
		\caption{Multi-class ROC curves (one-vs-rest) for the top-10 classes, with AUC values in the legend.}
		\label{fig:roc_curves}
	\end{figure}

\begin{figure}[htb]
		\centering
		\includegraphics[width=0.9\columnwidth]{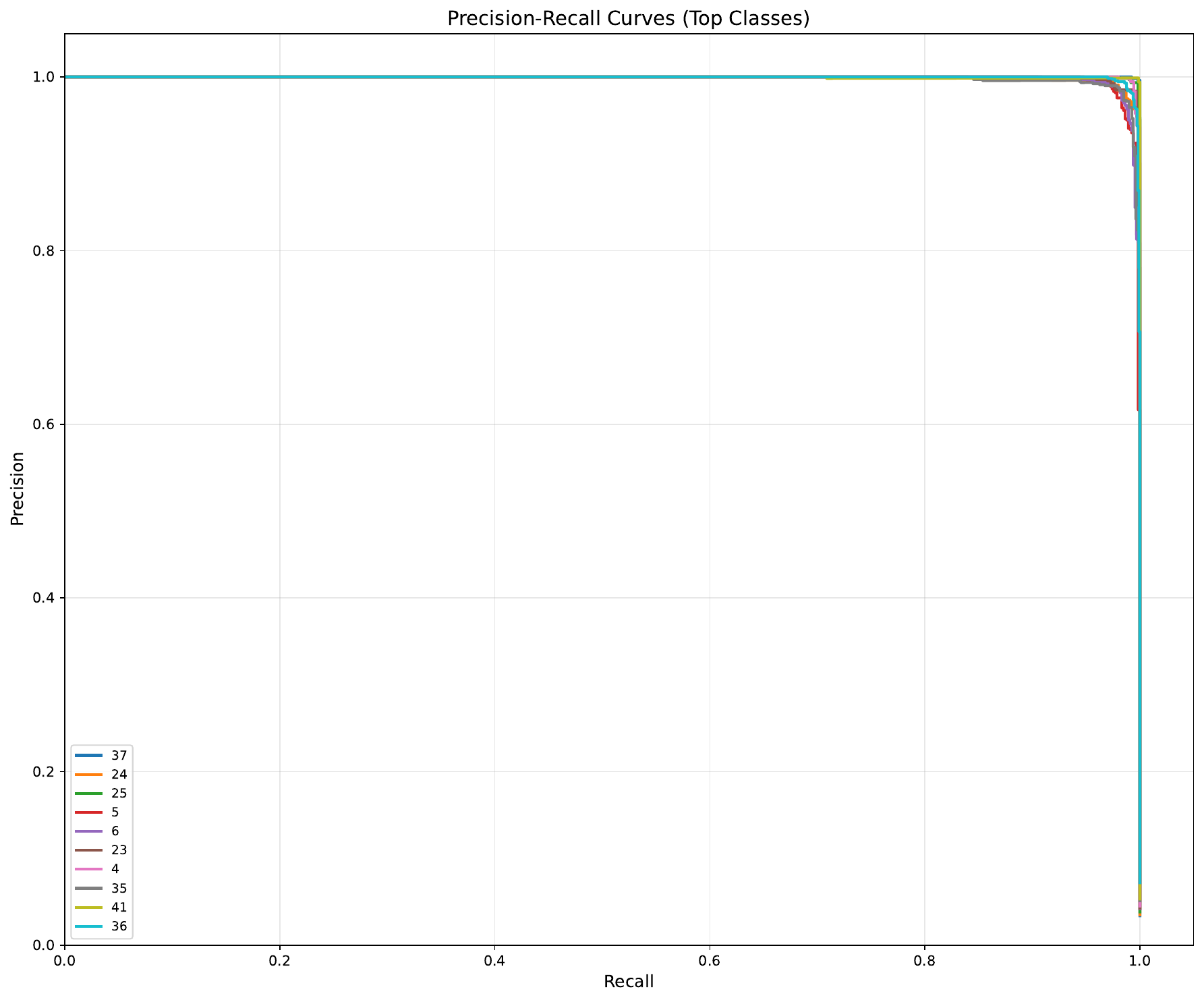}
		\caption{Precision--recall curves for the top-10 classes, demonstrating high precision across recall levels.}
		\label{fig:pr_curves}
	\end{figure}

	To characterize class-wise discriminability independent of a single operating threshold, Fig.~\ref{fig:roc_curves} presents multi-class, one-vs-rest ROC curves for the ten most frequent classes, with the corresponding area-under-curve (AUC) values given in the legend. The precision--recall curves in Fig.~\ref{fig:pr_curves} show that precision remains high even at elevated recall, a property of particular importance for imbalanced settings in which the correct identification of rare classes carries disproportionate operational significance. 
	
\begin{figure}[htb]
		\centering
		\includegraphics[width=0.9\columnwidth]{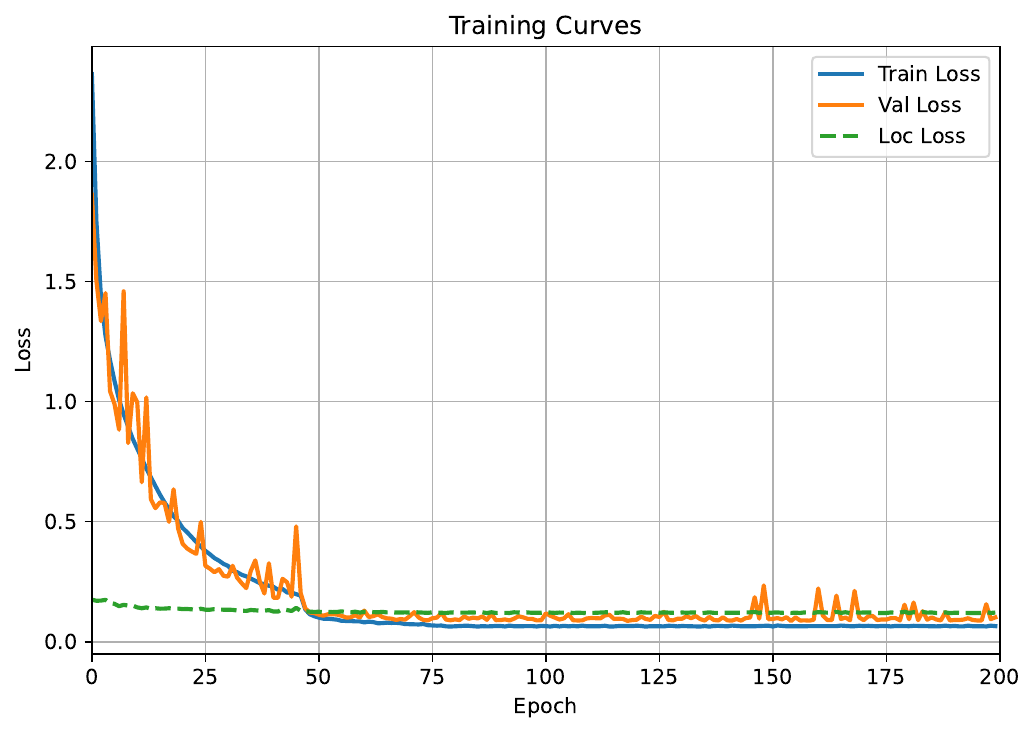}
		\caption{Training, validation, and localization loss curves over 200 epochs, showing stable convergence with occasional transient validation-loss spikes.}
		\label{fig:loss_curves}
	\end{figure}

\begin{figure}[htb]
		\centering
		\includegraphics[width=0.9\columnwidth]{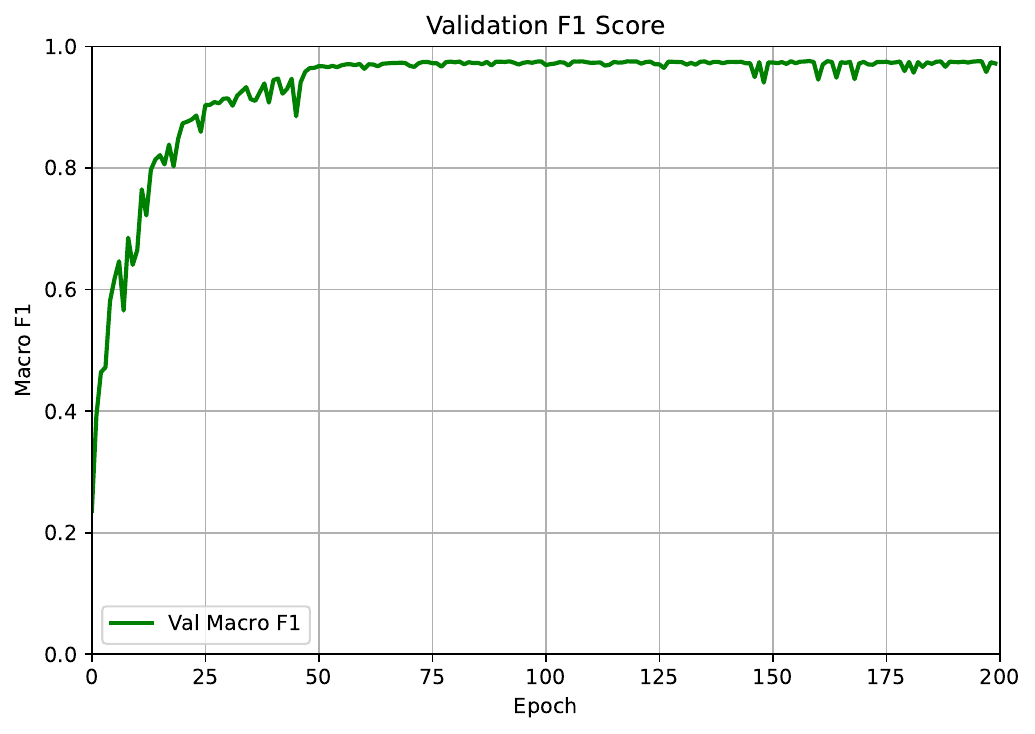}
		\caption{Validation macro F1-score progression from 23.60\% (epoch 1) to a peak of 97.61\% (epoch 159).}
		\label{fig:f1_curve}
	\end{figure}

	Training dynamics are illustrated in Figs.~\ref{fig:loss_curves} and \ref{fig:f1_curve}. As shown in Fig.~\ref{fig:loss_curves}, the training loss decreases from 2.3631 at epoch 1 to 0.0648 at epoch 200 and plateaus at roughly 0.063--0.068 after about epoch 80. The validation loss falls from 1.8607 to a minimum of 0.0859 (epoch 155) and typically fluctuates between 0.086 and 0.12 over the last 100 epochs; the gap to the training loss is small (about 0.02--0.03), consistent with mild generalization slack rather than overfitting. The validation loss also shows isolated transient spikes (up to 0.2345 at epoch 149), during which the validation macro F1-score briefly dips to as low as 0.9406 before recovering at the next epoch; these excursions do not affect the reported results, which use the best-F1 checkpoint. The auxiliary localization loss decreases from 0.1748 to about 0.121 (0.1213 at epoch 200) and stays within 0.119--0.125 over the last 100 epochs, indicating that the composite objective in Eq.~\eqref{eq:loss} is optimized stably despite coupling two distinct supervisory signals. Figure~\ref{fig:f1_curve} tracks the validation macro F1-score, which rises from 0.2360 at epoch 1 to above 0.97 by about epoch 66, stabilizes around 0.974--0.976, and peaks at 0.9761 at epoch 159. The final macro F1-score on the test partition, 0.9863, is computed on the full held-out set using this checkpoint. 
	
	\subsection{Localization Performance}
	\label{subsec:localization_results}
	
	Beyond event classification, the localization head identifies which of the four principal relays (R1--R4) are implicated in a detected event. Table~\ref{tab:localization_accuracy} reports per-relay binary accuracy: all four relays exceed 95.5\%, with an average accuracy of 95.93\%. This indicates that PTMTD can recover the Z-score-defined notion of the electrical locus of a disturbance (Section~\ref{subsec:loc_targets}), subject to two caveats. First, these targets are themselves derived from measurement statistics rather than independently verified field labels. Second, the targets are sparse (only 4.4--5.1\% of attack-labeled samples are positive for any given relay), so per-relay accuracy is dominated by the negative class, and a trivial all-negative predictor would already score close to 95\% on such samples. Binary accuracy therefore overstates localization skill and should be read as a coarse indicator; per-relay precision and recall on the positive class are the more informative measures. 
	
	\begin{table}[h]
		\centering
		\caption{Relay-wise localization accuracy (against Z-score-derived targets).}
		\label{tab:localization_accuracy}
		\begin{tabular}{lc}
			\toprule
			\textbf{Relay} & \textbf{Accuracy} \\
			\midrule
			R1 & 96.04\% \\
			R2 & 95.57\% \\
			R3 & 95.85\% \\
			R4 & 96.27\% \\
			\midrule
			\textbf{Average} & \textbf{95.93\%} \\
			\bottomrule
		\end{tabular}
	\end{table}
	
\begin{figure}[htb]
		\centering
		\includegraphics[width=0.9\columnwidth]{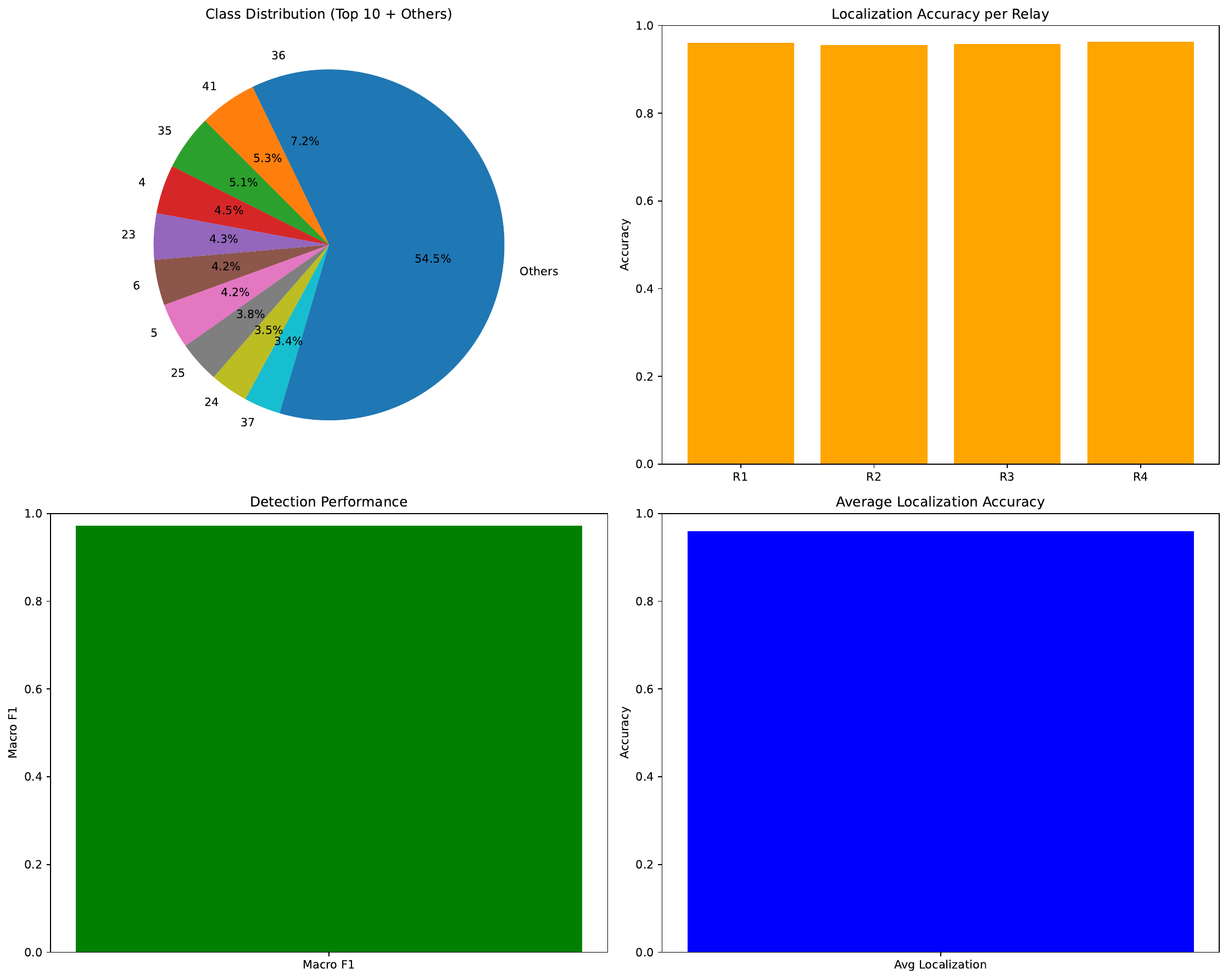}
		\caption{System overview: (a) class distribution (top-10 classes plus ``Others''), (b) relay abnormality rates during attacks, (c) detection macro F1-score, and (d) average localization accuracy.}
		\label{fig:system_overview}
	\end{figure}

	Figure~\ref{fig:system_overview} synthesizes four complementary perspectives on system behavior: (a) the class-distribution profile for the ten most frequent classes, with the remainder aggregated as ``Others,'' illustrating the underlying class imbalance; (b) relay-wise abnormality rates during attack conditions (4.4\%, 5.1\%, 5.0\%, and 4.4\% for R1--R4), which are of similar magnitude across relays, indicating that no single relay dominates the localization targets of Eq.~\eqref{eq:loc_target}; (c) the detection macro F1-score of 98.63\%; and (d) the average localization accuracy of 95.93\%. Taken together, these panels depict PTMTD's end-to-end behavior, from multi-class attack detection through relay attribution, within a single framework.
	
	\subsection{Comparison with State-of-the-Art Baselines}
	\label{subsec:comparison}
	
	Table~\ref{tab:baseline_comparison} situates PTMTD relative to GraphKAN, LSTM, GCN, and Random Forest. PTMTD achieves the highest accuracy and macro F1-score among all compared methods. In accuracy, it outperforms GraphKAN by 5.07 percentage points, LSTM by 6.28, GCN by 8.48, and Random Forest by 9.38; in macro F1-score, the corresponding margins are 6.27, 7.48, 9.58, and 10.53 points. 
	Paired $t$-tests across five independent runs with distinct random seeds confirm that these improvements are statistically significant ($p < 0.05$); we note that statistical significance at this sample size does not by itself establish practical significance, and the 5.07-point accuracy margin over GraphKAN in particular should be read as a modest, though consistent, improvement rather than a decisive one.
	
	\begin{table}[h]
		\centering
		\caption{Comparison of PTMTD with baseline methods.}
		\label{tab:baseline_comparison}
		\begin{tabular}{lcc}
			\toprule
			\textbf{Model} & \textbf{Accuracy (\%)} & \textbf{Macro F1 (\%)} \\
			\midrule
			PTMTD (Ours)              & \textbf{98.70} & \textbf{98.63} \\
			GraphKAN~\cite{wu2025graph} & 97 & 97 \\
			LSTM             & 92.42 & 91.15 \\
			GCN             & 90.22 & 89.05 \\
			Random Forest (500 trees) & 89.32 & 88.10 \\
			\bottomrule
		\end{tabular}
	\end{table}
	
	We attribute PTMTD's advantage over these baselines to three complementary factors. First, the dilated TCN extends the effective temporal receptive field without a proportional growth in parameters, which is particularly consequential for stealthy attacks whose signatures accrue gradually across several sampling intervals rather than appearing as an abrupt discontinuity. Second, the GATv2 attention mechanism reweights relay interactions dynamically rather than statically, allowing the model to concentrate representational capacity on the relays most implicated in an unfolding event. Third, the joint optimization of detection and localization under the composite objective of Eq.~\eqref{eq:loss} regularizes the shared representation toward features that are simultaneously discriminative and spatially informative, an effect that the ablation results below quantify directly. Beyond the aggregate margin over GraphKAN, we consider PTMTD's simultaneous relay-level localization -- a capability GraphKAN does not provide -- to be at least as consequential a contribution as the accuracy gain itself.
	
	\subsection{Ablation Studies}
	\label{subsec:ablation}
	
	To isolate the contribution of each architectural component, we conduct controlled ablations in which a single module is removed or substituted while all other hyperparameters are held fixed (Table~\ref{tab:ablation}). 
	
	\begin{table}[htbp]
		\centering
		\caption{Ablation study: impact of each component on detection accuracy and macro F1. $\Delta$ denotes the drop relative to the full model (percentage points).}
		\label{tab:ablation}
		
		\begin{tabularx}{\columnwidth}{>{\raggedright\arraybackslash}Xcccc}
			\toprule
			\textbf{Model Variant} & \textbf{Acc. (\%)} & \textbf{$\Delta$Acc.} & \textbf{Macro F1 (\%)} & \textbf{$\Delta$F1} \\
			\midrule
			PTMTD (Full)                                & \textbf{98.70} & --   & \textbf{98.63} & --   \\
			Without GAT (GCN)                           & 91.83 & 6.87 & 90.67 & 7.96 \\
			Without Temporal Encoding                   & 92.14 & 6.56 & 90.89 & 7.74 \\
			Without Residual Connections                & 92.45 & 6.25 & 91.02 & 7.61 \\
			Without Z-score Localization Supervision    & 93.12 & 5.58 & 91.85 & 6.78 \\
			\bottomrule
		\end{tabularx}
	\end{table}
	
	Replacing GATv2 with standard graph convolution reduces accuracy by 6.87 points (to 91.83\%) and macro F1 by 7.96 points, the largest degradation among the four ablations, and thereby identifies attention-based aggregation as the single most consequential architectural choice in PTMTD. Removing the temporal encoder -- feeding raw window features directly to the graph module -- lowers accuracy by 6.56 points (to 92.14\%) and macro F1 by 7.74 points, confirming that temporal context extracted prior to graph propagation is not redundant with the spatial reasoning performed downstream. Eliminating the residual connections and layer normalization surrounding the GATv2 stack yields a 6.25-point accuracy decline (to 92.45\%) and a 7.61-point macro F1 decline, consistent with the well-established tendency of deep graph networks toward over-smoothing in the absence of such stabilizing pathways. Finally, substituting the statistically derived Z-score localization targets with a naively supervised binary classifier lowers detection accuracy to 93.12\% (a 5.58-point decline) and localization accuracy to 95.2\% (compared with 95.93\% for the full model; reported in the text only), indicating that the benefit of statistically grounded supervision extends beyond the localization head to the shared representation that both tasks draw upon. 
	
	Removing any single component costs between 5.58 and 6.87 points of accuracy, so all four contribute materially. Because only single-component removals were evaluated, these drops are not additive, and potential interactions between components are not isolated by this design.
	
	\section{Discussion}
	\label{sec:discussion}
	
	\subsection{Interpretation of Results}
	\label{subsec:interpretation}
	
	The reported results indicate that PTMTD addresses the dual challenge of cyber-attack detection and spatial localization in power system protection networks within a single, jointly trained framework. The detection accuracy of 98.70\% and macro F1-score of 98.63\% across 37 event classes -- including subtle cyber-attacks that conventional threshold-based or shallow statistical detectors are prone to miss -- suggest that the combination of temporal and relational modeling captures discriminative structure that either component would fail to expose in isolation, as the ablations of Section~\ref{subsec:ablation} indicate.
	
	The localization result (95.93\% average per-relay accuracy) is operationally attractive, since accurate spatial attribution would enable faster isolation and mitigation of an ongoing disturbance, which is directly relevant to preventing cascading failures in interconnected power networks. That this accuracy is obtained with automatically generated, statistically derived labels -- rather than costly manual annotation -- suggests a practical path toward deploying similar localization capabilities where labeled ground truth is otherwise unavailable. We stress, however, two limits on its interpretation: the figure measures agreement with a Z-score-derived proxy rather than independently verified ground truth (Section~\ref{subsec:loc_targets}), and, because the positive targets are sparse (4.4--5.1\% of attack-labeled samples), binary accuracy is only modestly informative on its own (Section~\ref{subsec:localization_results}).
	
	Finally, the model's modest footprint (639{,}177 parameters, 2.72 MB) and low estimated inference latency (approximately 2.5 ms per sample on CPU, based on preliminary profiling) indicate that the accuracy gains reported here need not come at the cost of deployability. This combination of accuracy and efficiency is a prerequisite, though not by itself sufficient, for real-time deployment on the resource-constrained embedded platforms typical of protective-relay hardware.
	
	\subsection{Limitations and Future Work}
	\label{subsec:limitations_future}
	
	Several limitations qualify the scope of these findings and motivate future work.
	
	\paragraph{Scope of the evaluation dataset.} The MSU--ORNL benchmark used throughout this paper is a transmission-level relay-protection testbed; it does not include DER, inverter-based generation, or islanding dynamics characteristic of true microgrids. While the PTMTD architecture makes no assumption specific to transmission-level topology -- it operates purely on relay measurement correlations and generic temporal windows -- we have not validated its performance on a DER-integrated microgrid dataset, and we do not claim microgrid-specific results here. Establishing this generalization, ideally on a dataset such as those used in~\cite{ogiesobaeguakun2026cyberattackdetectionvirtualizedmicrogrids} or a hardware-in-the-loop DER testbed, is a priority for follow-up work.
	
	\paragraph{Circularity and sparsity of localization labels.} The localization targets in Eq.~\eqref{eq:loc_target} are derived from the same measurement channels supplied to the model as input. Consequently, the 95.93\% localization accuracy reported in Section~\ref{subsec:localization_results} demonstrates that PTMTD can recover this specific, statistically defined notion of relay involvement; it does not by itself demonstrate agreement with independently verified, field-confirmed fault locations. In addition, positive targets are rare (4.4--5.1\% of attack-labeled samples), so accuracy is dominated by the negative class; future reporting should include per-relay precision, recall, F1, and precision--recall AUC, and comparison against manually annotated or physically instrumented ground truth where available.
	
	\paragraph{Evaluation protocol.} Because the windows overlap with stride one, neighboring windows share 19 of their 20 time steps. A stricter test of generalization would hold out entire operational scenarios (e.g., training on 12 scenario files and testing on the remaining 3), and k-fold or multiple-partition cross-validation would further strengthen confidence in the reported margins, particularly the 5.07-point accuracy margin over GraphKAN. The present evaluation relies on a single train/test partition (repeated only across random seeds).
	
	\paragraph{Other limitations.} The graph in Section~\ref{subsec:graph_construction} is constructed from static, dataset-wide correlations and does not adapt to transient topology changes arising from network reconfiguration or islanding; dynamic graph-learning methods that update the adjacency matrix from streaming measurements represent a natural extension. The current formulation assumes a fixed relay count ($N = 4$); generalizing to larger networks with a variable number of substations will likely require node-wise pooling or attention-based aggregation mechanisms that do not presuppose a fixed graph size. The temporal window length ($T = 20$) was selected empirically rather than through systematic search; a sensitivity analysis spanning multiple window sizes (e.g., 10, 20, 30, 40, 50) is needed to characterize the accuracy--latency trade-off more precisely. The less frequent classes (e.g., classes 22 and 27) show the lowest per-class scores, and class-specific augmentation or re-weighting could help. Finally, the Z-score criterion of Eq.~\eqref{eq:loc_target}, while statistically motivated, may not capture all classes of physical anomaly -- subtle power swings, for instance, may not manifest as large deviations in mean relay measurements; richer physics-informed criteria, such as impedance-based or power-flow residual violations, could plausibly sharpen localization further. Looking beyond detection and localization, a natural extension is to close the loop with active defense: using the localization output to trigger automated reconfiguration -- for example, isolating an implicated relay -- so that detection, attribution, and mitigation form a single automated pipeline.
	
	\section{Conclusion}
	\label{sec:conclusion}
	
	We have presented PTMTD, a Pearson-driven, multimodal temporal-graph architecture for the joint detection and relay-level localization of cyber-attacks in power system protection networks. By coupling a dilated TCN with a residually stabilized, multi-head GATv2 module and a dual-head output structure, PTMTD learns temporal and relational representations jointly optimized for both classification and spatial attribution. The underlying graph is inferred directly from measurement correlations, avoiding dependence on explicit impedance data, while a composite loss ties detection and localization together within a single training objective. On the MSU--ORNL transmission-protection benchmark -- 78{,}377 samples spanning 15 operational scenarios and 37 event classes -- PTMTD attains 98.70\% detection accuracy, 98.63\% macro F1-score, and 95.93\% average relay-wise localization accuracy (the latter measured against statistically derived labels rather than independently verified ground truth). These results exceed GraphKAN (93.63\% accuracy), LSTM (92.42\%), GCN (90.22\%), and Random Forest (89.32\%) by margins of 5.07 to 9.38 percentage points in accuracy and 6.27 to 10.53 points in macro F1-score, with statistical significance ($p < 0.05$) assessed across independently seeded runs. Ablation analysis shows that removing temporal encoding, graph attention, residual stabilization, or Z-score-derived localization supervision lowers accuracy by 6.56, 6.87, 6.25, and 5.58 percentage points, respectively -- supporting the architecture's design rationale.
	
	The normalized confusion matrix and per-class analysis show that every one of the 37 classes attains recall of at least 0.94, with the lowest scores on the least represented classes, while the ROC and precision--recall curves corroborate the model's discriminative strength. Combined with a compact parameter footprint (639{,}177 parameters, 2.72 MB) and low estimated inference latency (approximately 2.5 ms per sample on CPU), PTMTD's accuracy is achieved without sacrificing the efficiency required for real-time operation on constrained embedded hardware. Its relay-level localization (95.93\% average accuracy against Z-score-derived targets) positions the framework as a promising basis for rapid fault isolation -- a capability central to preventing cascading failures in interconnected power networks, including but not limited to microgrids -- pending the field validation and the more discriminating localization metrics discussed in Section~\ref{subsec:limitations_future}.
	
	Future work will pursue five directions: dynamic graph construction that adapts to online topology changes; architectural extensions supporting a variable number of relay nodes; the incorporation of richer physics-informed constraints, such as power-flow residual violations, into the localization objective; validation on DER-integrated microgrid and other field-collected operational data, together with scenario-wise hold-out evaluation, to establish the generalization only conjectured here; and the integration of active-defense mechanisms that automatically respond to detected and localized attacks, closing the loop between detection and mitigation. We view PTMTD as a step toward interpretable, efficient, and domain-knowledge grounded learning-based protection for cyber-physical power systems, and we release our code and pre-trained models to support reproducibility and further research in this area.
	
	\bibliographystyle{plain}
	\bibliography{DataDrivenFeedBack}

\begin{thebibliography}{10}

\bibitem{11098679}
Zulfiqar Ali, Tahir Hussain, Chun-Lien Su, Irfan Khan, Anca~Delia Jurcut,
  Shao-Hang Tsao, Cho-Han Hu, and Mahmoud Elsisi.
\newblock Deep learning-driven cyber attack detection framework in dc shipboard
  microgrids system for enhancing maritime transportation security.
\newblock {\em IEEE Transactions on Intelligent Transportation Systems},
  26(11):20122--20142, 2025.

\bibitem{ali2025cyber}
Zulfiqar Ali, Chun-Lien Su, Yacine Terriche, Seyed~Hossein Rouhani,
  Le~Quang~Nhat Hoang, Muhammad Sadiq, Shao-Hang Tsao, Syed~Zagam Abbas, Ejaz
  Ahmad, and Mahmoud Elsisi.
\newblock Cyber resilience in shipboard microgrids: adaptive hybrid artificial
  intelligent methods and systematic review.
\newblock {\em Neural Computing and Applications}, 37(22):17633--17674, 2025.

\bibitem{BALAHEWA2026113148}
Imasha Balahewa, Lars Bjorndal, Chris Mi, and Tong Huang.
\newblock Cyber attack detection in renewable-rich ibr-dominated microgrids: A
  natural watermarking approach.
\newblock {\em Electric Power Systems Research}, 260:113148, 2026.

\bibitem{BARBHAYA2025109278}
Malhar Barbhaya, Purushottama~Rao Dasari, Seshu~Kumar Damarla, Rajagopalan
  Srinivasan, and Biao Huang.
\newblock A deep learning framework for cyberattack detection and
  classification in industrial control systems.
\newblock {\em Computers \& Chemical Engineering}, 202:109278, 2025.

\bibitem{11079664}
Fabin Cheng, Jingang Lai, and Zhigang Zeng.
\newblock Bilevel distributed secure control of dc microgrid clusters under
  scale random cyberattacks.
\newblock {\em IEEE Transactions on Smart Grid}, 16(6):4324--4334, 2025.

\bibitem{eswaran2025cyberattack}
Nanthaluxsan Eswaran, Jalini Sivarajah, Kopikanth Karunakaran, Logeeshan
  Velmanickam, Sisil Kumarawadu, and Chathura Wanigasekara.
\newblock Cyberattack detection and classification of power converters in
  islanded microgrids using deep learning approaches.
\newblock {\em Electronics}, 14(17):3409, 2025.

\bibitem{fakhrooeian2025artificial}
Mahan Fakhrooeian, Ali Basem, Mohammad~Mahdi Gholami, Nahal Iliaee,
  Alireza~Mohammadi Amidi, Amin~Heydarian Hamzehkanloo, and Akbar Karimipouya.
\newblock An artificial insurance framework for a hydrogen-based microgrid to
  detect the advanced cyberattack model.
\newblock {\em Scientific Reports}, 15(1):3762, 2025.

\bibitem{ghiasi2021cyber}
Mohammad Ghiasi, Moslem Dehghani, Taher Niknam, Abdollah Kavousi-Fard,
  Pierluigi Siano, and Hassan~Haes Alhelou.
\newblock Cyber-attack detection and cyber-security enhancement in smart
  dc-microgrid based on blockchain technology and hilbert huang transform.
\newblock {\em Ieee Access}, 9:29429--29440, 2021.

\bibitem{HEMMATI2026120728}
Reza Hemmati and Hedayat Saboori.
\newblock A hybrid data-driven and physics-informed energy management model for
  electrical grids with spatio-temporal cyberattack detection-reconstruction
  using digital twin.
\newblock {\em Energy Conversion and Management}, 348:120728, 2026.

\bibitem{hosseini2026cyber}
Seyyed~Mohammad Hosseini~Rostami, Mahdi Pourgholi, and Hadi Asharioun.
\newblock Cyber attack and fault detection in dc microgrids by designing an
  event-triggered based-robust algorithm.
\newblock {\em Scientific Reports}, 16(1):1701, 2026.

\bibitem{jena2021cyber}
Satabdy Jena, Narayana~Prasad Padhy, and Josep~M Guerrero.
\newblock Cyber-resilient cooperative control of dc microgrid clusters.
\newblock {\em IEEE systems Journal}, 16(2):1996--2007, 2021.

\bibitem{11168950}
Mohit Kachhwaha and Deepak Fulwani.
\newblock Robust covert attack detection in dc microgrids: A novel approach to
  cybersecurity.
\newblock {\em IEEE Transactions on Industrial Electronics}, 73(1):447--457,
  2026.

\bibitem{ogiesobaeguakun2026cyberattackdetectionvirtualizedmicrogrids}
Osasumwen~Cedric Ogiesoba-Eguakun and Suman Rath.
\newblock Cyberattack detection in virtualized microgrids using lightgbm and
  knowledge-distilled classifiers, 2026.

\bibitem{orman2025cyberattack}
Abdullah Orman.
\newblock Cyberattack detection systems in industrial internet of things (iiot)
  networks in big data environments.
\newblock {\em Applied Sciences}, 15(6):3121, 2025.

\bibitem{RAGAB2025438}
Mahmoud Ragab, Mohammed Basheri, Nasser~N. Albogami, Alanoud Subahi, Omar~A.
  Abdulkader, Hashem Alaidaros, Hassan Mousa, and Abdullah AL-Malaise
  AL-Ghamdi.
\newblock Artificial intelligence driven cyberattack detection system using
  integration of deep belief network with convolution neural network on
  industrial iot.
\newblock {\em Alexandria Engineering Journal}, 110:438--450, 2025.

\bibitem{razani2026robust}
Shahin Razani, Mahsan Tavakoli-Kakhki, and Ahmad Kalhor.
\newblock Robust data-driven synchronization of heterogeneous nonlinear
  multi-agent systems.
\newblock {\em Nonlinear Dynamics}, 114(5):372, 2026.

\bibitem{razani2025indirect}
Shahin Razani, Majid Zarei, and SM~Mehdi Dehghan.
\newblock Indirect data-driven resilient control against bias attacks and
  disturbance using sparse identification of nonlinear dynamics.
\newblock {\em IEEE Access}, 2025.

\bibitem{saqib}
Nazmus Saqib, Souradeep Bhattacharya, Kiran~Kumar Challa, and Manimaran
  Govindarasu.
\newblock Attack resilient microgrid: Impact characterisation, online anomaly
  detection and mitigation.
\newblock {\em IET Smart Grid}, 9(1):e70070, 2026.

\bibitem{SUBRAMANIAM2025103400}
Revathi Subramaniam, Androse~Joseph Sheela, and Abdullah Alwabli.
\newblock Enhanced cybersecurity and cyber-attack detection in smart dc micro
  grids using blockchain technology and svm technique.
\newblock {\em Ain Shams Engineering Journal}, 16(7):103400, 2025.

\bibitem{11015976}
Koduru~Sriranga Suprabhath, Venkata Siva Prasad~Machina, Sreedhar Madichetty,
  and Sukumar Mishra.
\newblock Securing dc microgrids against cyberattacks: Hybrid physics-informed
  neural network control strategy with real-time implementation.
\newblock {\em IEEE Journal of Emerging and Selected Topics in Power
  Electronics}, 13(5):6292--6304, 2025.

\bibitem{tan2025developments}
Sen Tan, Peilin Xie, Juan~C Vasquez, and Josep~M Guerrero.
\newblock Developments, challenges and future opportunities in cybersecure
  microgrid control.
\newblock {\em Nature Reviews Electrical Engineering}, 2(8):522--540, 2025.

\bibitem{wu2025graph}
Ying Wu, Zhiyuan Zang, Xitao Zou, Wentao Luo, Ning Bai, Yi~Xiang, Weiwei Li,
  and Wei Dong.
\newblock Graph attention and kolmogorov--arnold network based smart grids
  intrusion detection.
\newblock {\em Scientific Reports}, 15(1):8648, 2025.

\bibitem{ZHENG2026112529}
Haofeng Zheng, Xingmei Li, and Fengyun Li.
\newblock Risk-aware unsupervised cyberattack detection via contrastive
  autoencoding and adversarial training in smart grid substations.
\newblock {\em Reliability Engineering \& System Safety}, 272:112529, 2026.

\end{thebibliography}
	
\end{document}